\documentclass{ws-ijmpc}

\usepackage{hyperref}
\usepackage{graphicx}
\usepackage{amssymb}
\usepackage{amsmath}
\renewcommand\draftnote{}

\newcommand{\bea}{\begin{eqnarray}}
\newcommand{\eea}{\end{eqnarray}}

\begin{document}

\markboth{A.V.~Bednyakov and \c{S}.H.~Tany\i ld\i z\i}{}

\catchline{}{}{}{}{}

\title{A {\tt MATHEMATICA} PACKAGE FOR CALCULATION OF ONE-LOOP PENGUINS IN FCNC PROCESSES
}

\author{Alexander Vadimovich Bednyakov}

\address{Bogoliubov Laboratory of Theoretical Physics, Joint
Institute for Nuclear Research, 141980, Dubna, Moscow Region\\
bednya@theor.jinr.ru}

\author{\c{S}\"ukr\"u Hanif Tany\i ld\i z\i}

\address{Bogoliubov Laboratory of Theoretical Physics, Joint
Institute for Nuclear Research, 141980, Dubna, Moscow Region\\
hanif@theor.jinr.ru}

\maketitle


\begin{abstract}
In this work, we present a {\tt Mathematica} package {\tt Peng4BSM@LO} which calculates the contributions to the Wilson Coefficients of certain effective operators originating from the one-loop penguin Feynman diagrams. Both vector and scalar external legs are considered.
The key feature of our package is the ability to find the corresponding expressions in almost any
New Physics model which extends the SM and has no flavour changing neutral current (FCNC) transitions
at the tree level.

\end{abstract}

\ccode{PACS Nos.: 11.25.Hf, 123.1K}

\section{Introduction}
\label{intro}

The flavour changing neutral current (FCNC) processes attract a lot of interest from both the theoretical and experimental side. Such transitions are absent in the SM at the tree level and, thus, are suppressed in comparison with the charged current processes. Due to this, FCNC can be used as an excellent probe of New Physics, which can considerably alter the predictions of the SM.  
At the moment, no significant deviations are found. Consequently, these rare processes impose very important constraints on Beyond-the-SM (BSM) physics. 
Typical examples are the $b \to s \gamma$~\cite{Lees:2012ym,Lees:2012ufa,Hermann:2012fc} and $B_s\to\mu\mu$ decays~\cite{Lees:2013kla,Aaij:2013aka,Chatrchyan:2013bka} which are used in different studies of various supersymmetric extensions of the SM. Since new particles predicted by BSM models are usually much heavier than the SM ones, the corresponding (short-distance) contribution to FCNC amplitudes can be absorbed into the Wilson coefficients of the operators which enter into the weak effective Hamiltonian \cite{Buras:1998raa}.

In order to calculate the Wilson coefficients for a particular FCNC process in supersymmetric or any two-higgs doublet models, one can use different codes available on the market, e.g., {\tt SuperIso}~\cite{Mahmoudi:2008tp,Mahmoudi:2009zz,Arbey:2011zz}, {\tt SUSY\_FLAVOUR}~\cite{Rosiek:2010ug,Crivellin:2012jv} or {\tt SPheno\_v3}~\cite{Porod:2011nf} (see also, Ref.~\cite{Mahmoudi:2010iz}). 

We created a {\tt Mathematica} package {\tt Peng4BSM@LO} which can be used together with FeynArts~\cite{Eck:dissertation,Hahn:2000kx,Hahn:2001rv,Fritzsche:2013fta,famanual} and FeynCalc
~\cite{mertig} to address the same problem. However, contrary to the above-mentioned flavour codes, our routines give an opportunity to obtain the expression for the Wilson coefficients in almost any renormalizable BSM model, 
which can be implemented in {\tt FeynArts} format with the help of {\tt FeynRules}~\cite{Christensen:2008py,Christensen:2009jx,Alloul:2013bka}, {\tt LanHEP}~\cite{Semenov:2008jy,Semenov:2010qt} or {\tt SARAH}~\cite{Staub:2008uz,Staub:2009bi,Staub:2012pb,Staub:2013tta}.
However, it should be stressed from the very beginning that a BSM model should \emph{not} have the tree-level FCNC coupling, which corresponds to the considered FCNC (sub)process.
Only one-loop generated transitions are taken into account.

In Sec.\ref{op}, we define our notation and present generic 
operators contributing to the effective Hamiltonian. In Sec.\ref{structure}, we introduce the programming instruments, our method to define and calculate the Wilson coefficients and describe and explain the structural diagram of the package {\tt Peng4BSM@LO}. In Sec.\ref{use}, the basic usage of the functions of {\tt Peng4BSM@LO} is presented. In Sec.\ref{test}, we carry out some benchmark tests of the 
package by reproducing known expressions for penguins with external $Z$-boson, photon $\gamma$ \cite{inamilim} and the Higgs field\footnote{In the initial version of the package only vector external operators were considered.} $H$ \cite{Dedes:2003kp} in the SM.
In addition, the gluino contribution in MSSM with non-minimal flavour violation is recalculated for $b\to s\gamma$ process \cite{Bertolini:1990if}. In \ref{summary}, we give characteristics of the package and the reference, from which one may download it. In \ref{procedure}, commands' descriptions of the main procedures are stated. Then subsequently, in \ref{auxiliary}, descriptions of the auxiliary procedures and definitions of program parameters  are clarified.

\section{Generic operators}
\label{op}

In {\tt Peng4BSM@LO}, we consider the following generic effective local operators and their form factors, $\left(N^{0,c}_{a,b}\right)_{L,R}$, $\left(E^{0,c}_{a,b}\right)_{L,R}$,  $\left(E^{2,c}_{a,b}\right)_{L,R}$, $\left(M^{1,c}_{a,b}\right)_{L,R}$.
The scalar operators are of the form
\begin{eqnarray}
\label{s1}
H_{eff} \ni \left(\bar F_a^\prime P_{L,R}F_bS_c\right)\left(N^{0,c}_{a,b}\right)_{L,R}
\end{eqnarray}
where $S$ is a neutral scalar boson field and $P_{L,R}=(1/2)(1\mp\gamma^5)$ are  the projection operators. The monopole operators which conserve chirality are of the form
\begin{eqnarray}
\label{e0ande2}
H_{eff} \ni 
\left(\bar F_a^\prime \gamma^\mu P_{L,R}F_b\right)\left[\left(E^{0,c}_{a,b}\right)_{L,R}
g^{\mu\nu}
+ 
\left(g^{\mu\nu}q^2-q^\mu q^\nu\right)\left(E^{2,c}_{a,b}\right)_{L,R}
\right]V^c_\nu 
\end{eqnarray}
and the dipole operators which flip chirality are  
\begin{eqnarray}
\label{m1}
H_{eff} \ni
\left(\bar F_a^\prime \sigma^{\mu\nu} P_{L,R}F_bq_\mu V^c_\nu\right)\left(M^{1,c}_{a,b}\right)_{L,R}.
\end{eqnarray}
Here the metric tensor is defined as $g^{\mu\nu}=\mathrm{diag}(1,-1,-1,-1)$, $\sigma^{\mu\nu}\equiv (i/2)\left[\gamma^\mu,\gamma^\nu\right]$, and $q_\mu$ is the outgoing momentum of a neutral vector boson $V$ entering into the operators. Fermions of different 
families are denoted by $F$ and $F^\prime$, $a,~b,~c$ are some, e.g., color indices. From the form factors one can easily extract the corresponding Wilson coefficients. 
This kind of operators originates from the expansion of  penguin amplitudes in external momenta, so that 
($N^0$, $E^0$), $M^1$ and $E^2$ correspond to the zeroth, first and second order terms in this expansion, respectively.

\section{Structure of {\tt Peng4BSM@LO}}
\label{structure}

\begin{figure}[t!]
\begin{center}
\includegraphics[width=5in]{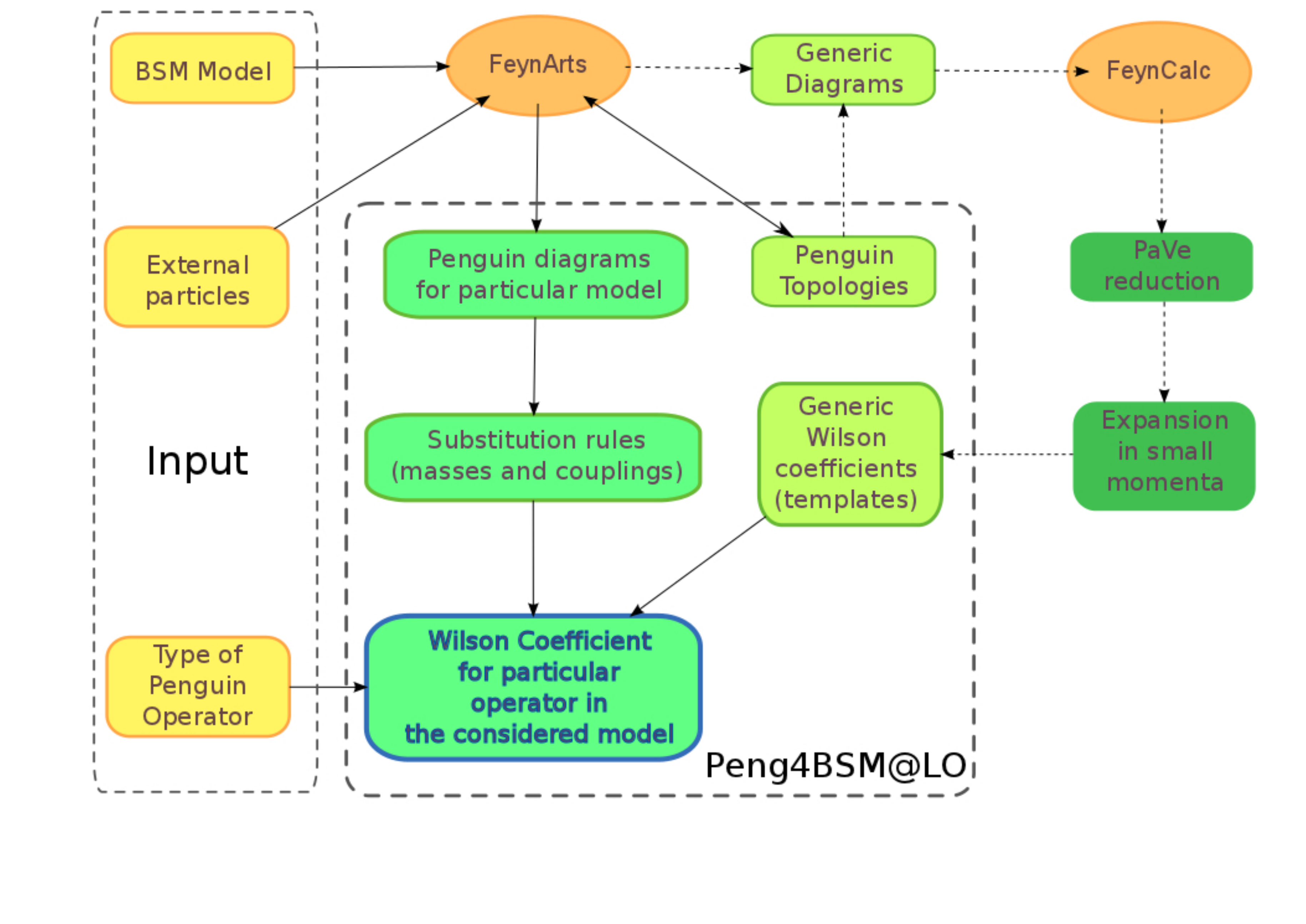}
\caption{The new package to calculate and define Wilson coefficients for the  one-loop penguin diagrams in FCNC processes. }
\label{pack}
\end{center}
\end{figure}

Our paradigm (see Fig.~\ref{pack}) is inspired by {\tt FeynArts} hierarchy of fields ({\tt Generic}, {\tt Classes}, and {\tt Particles}) and is  based on the fact that the Lorentz structure of Feynman vertices is fixed only by the type of  participating particles, so one can define and calculate generic Wilson coefficients originating from generic diagrams. 
The corresponding amplitudes involve generic couplings which can be substituted later by actual expressions. One should keep in mind that we heavily rely on the Lorentz structures defined in the generic model file {\tt Lorentz.gen} distributed with {\tt FeynArts}. The restriction applies primarily to the {\tt LanHEP} package which generates its own generic model file. Due to this, additional effort is required to rewrite Feynman rules produced by {\tt LanHEP} to make them consistent with our code.

\begin{figure}[t!]
\begin{center}
\includegraphics[width=5in]{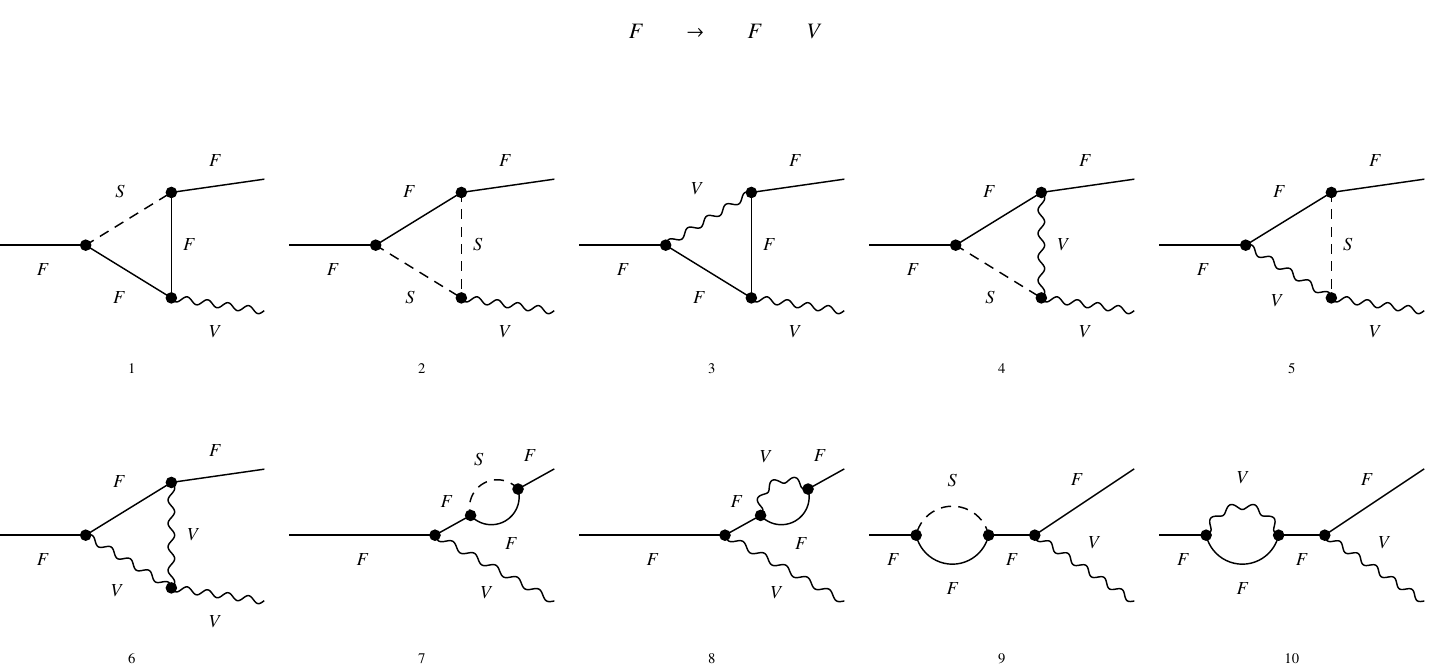}
\caption{The one-loop generic diagrams, where $F$ is a a fermion, $S$ is a scalar and $V$ is a vector. We do not consider the self-energy insertions in the neutral vector boson leg, since we assume that there is no tree-level FCNC. The corresponding diagrams for $F \to F~S$ transition are obtained by replacing the external vector boson $V$ with a scalar field $S$.}
\label{gen}
\end{center}
\end{figure}

With the help of {\tt FeynArts} we generate so-called generic diagrams (see Fig.~\ref{gen}). Then we reformulate the generic amplitudes by means of {\tt FeynCalc} in terms of the fundamental integrals of Passarino-Veltman's~\cite{passarino} which we expand in the limit of vanishing external momenta in order to obtain the generic Wilson coefficients. These generic Wilson coefficients serve as templates and can be used 
in any model. 
Given the corresponding expressions, the substitution rules for particular diagrams and amplitudes can be applied and the Wilson coefficients for a particular model can be obtained.

Let us mention that we restrict the package to the Feynman gauge, in which gauge propagators have very simple structure and, as a consequence, the complexity of the templates is significantly reduced. In addition, no unphysical gauge-parameter dependent masses appear in individual diagrams. 

\begin{figure}[t]
\begin{center}
\includegraphics[width=4in]{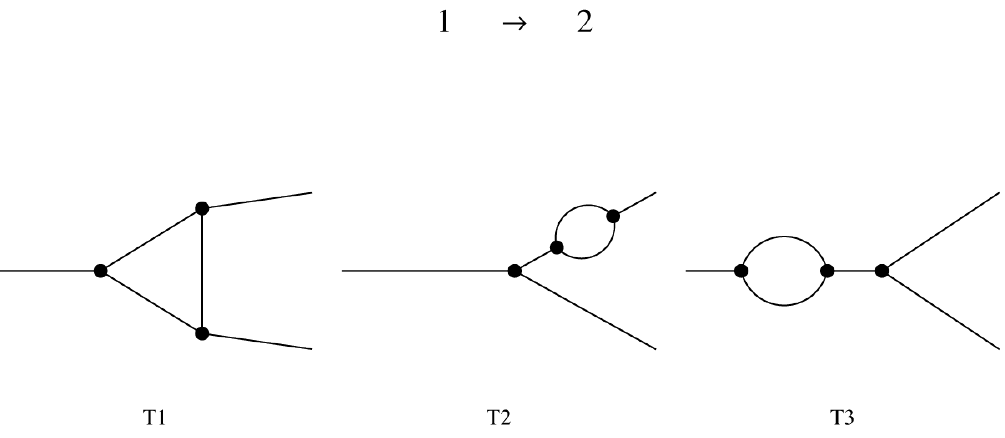}
\caption{Penguin topologies as defined in {\tt Peng4BSM@LO}. }
\label{pengtop}
\end{center}
\end{figure}

As we see from Fig.~\ref{pack}, in order to obtain a contribution to $N^0$, $E^{0,2}$ or $M^1$, one should define the model and the corresponding operator with external particles $F_a, F_b$, $V_c$ or $S_c$  from Eqs.~\eqref{s1},\eqref{e0ande2}, and \eqref{m1}. Given this input, the package can be used in the following way. First of all, one should construct  the relevant Feynman diagrams by calling {\tt PengInsertFields [$\{F_b\}$ $\to \{F_a, B_c\}, {\tt Model}\to \mbox{{\tt MOD}} ]$} where the boson $B_c$,  is either a vector one, $V_c$, or a scalar one denoted by $S_c$. 
The procedure is similar to {\tt InsertFields} of {\tt FeynArts} but uses a predefined set of topologies {\tt PenguinTopologies}\footnote{The absence of self-energy insertion in the third (vector/scalar) external line reflects the fact that there is no FCNC at the tree-level.} (see Fig.~\ref{pengtop}). 
The  model is specified by the option {\tt Model} $\to$ {\tt MOD}.  The corresponding Feynman rules in {\tt FeynArts} notation are taken from the file {\tt MOD.mod}. The function {\tt PengInsertFields} accepts the same options as {\tt InsertFields}, so one can restrict the set of generated diagrams, e.g., by {\tt ExcludeParticles} option\cite{famanual} . Then, the penguin amplitudes are produced by {\tt PengCreateFeynAmp}\footnote{Similar to CreateFeynAmp[dia] of {\tt FeynArts}, where dia is a list of diagrams.} from the diagrams created with {\tt PengInsertFields}. After that, one should apply {\tt ExtractPenguinSubsRules} to produce a list of substitution rules for each diagram (amplitude) generated by {\tt PengCreateFeynAmp}. The rules specify the actual couplings for each generic diagram (amplitude). They can be used later together with {\tt SubstituteMassesAndFeynmanRules} to obtain final analytic expressions in the considered model. In order to demonstrate the features of the package, we apply it to the study of the effective $d\bar s Z$, $d\bar s\gamma$~\cite{inamilim} and  $b\bar b H$~\cite{Dedes:2003kp} vertices in the SM, and to the effective $b\bar s\gamma$ ~\cite{Bertolini:1990if} vertex in the MSSM.

\section{{\tt Peng4BSM@LO} in Use}
\label{use}

The package {\tt Peng4BSM@LO} employs {\tt FeynArts} to generate the relevant diagrams. 
It is worth mentioning that one is not forced to use {\tt FeynCalc} which was utilized by us at the intermediate stages for calculation of the generic amplitudes. However, 
it may be convenient to load {\tt FeynCalc}\footnote{ {\tt FeynCalc} includes a patched version of {\tt FeynArts} in the distribution.} to produce some nicely formatted output.  In order to do so, one executes the following commands in the {\tt Mathematica} FrontEnd
\bea
\tt In[1]:=&&\tt \$UseFeynCalc = True;
\eea
To use the package {\tt Peng4BSM@LO} in the notebook, we should load it by the following command line
\bea
\tt In[2]:=\hspace{.7cm}\tt{Get}["/\tt{PATH}/\tt{Peng4BSMatLO.m}"];
\eea
where {\tt PATH} is the path to the directory with {\tt Peng4BSM@LO}. After this preparation, we can specify the operator external particles in terms of {\tt FeynArts} fields defined in the considered model 
\bea
\tt In[3]:=&&\tt{InF = F[4, \{1, c1\}];}~~~(\textrm{for d-quark})\nonumber\\
&&\tt{OutF = F[4, \{2, c2\}];}~~~(\textrm{for s-quark})\nonumber\\
&&\tt{OutV = V[2]};~~~(\textrm{for Z-boson})
\eea 

\begin{figure}[htb]
\begin{center}
\includegraphics[width=5in]{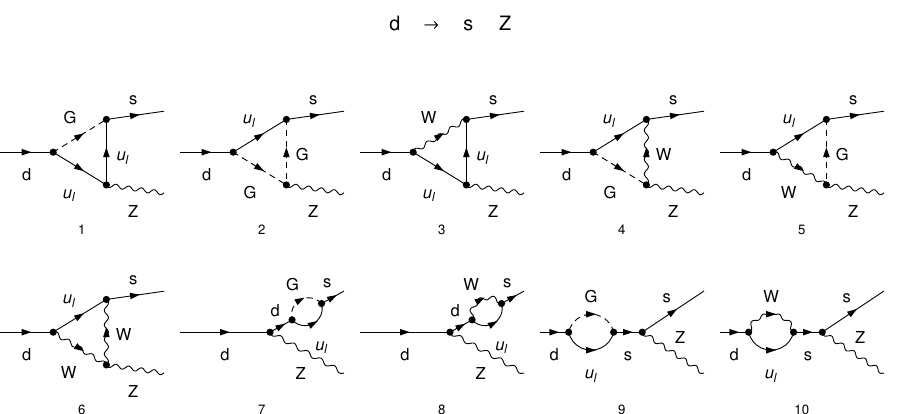}
\caption{The one-loop diagrams contributing to $d\bar s Z$ in the SM. The fields $u_l$, $W$, and $G$ in internal lines correspond to up-type quarks of the $l$-th generation,
$W$-boson and charged would-be Goldstone boson, respectively.}
\label{dsZSM}
\end{center}
\end{figure}

The next step is to distribute the fields defined in the chosen model over internal lines in the considered topologies. This is done automatically  by {\tt FeynArts} with the help of {\tt PengInsertFields} function,
\bea
\label{code5}
\tt In[4]:=&&\tt{diagrams = PengInsertFields[}\{\tt{InF}\}\nonumber\\
&&\to \{\tt{OutF}, \tt{OutV}\},\tt{Model} \to \tt{"SMQCD"]}\nonumber\\
\tt In[5]:=&&\tt{Paint[diagrams,PaintLevel} \to \tt{\{Classes\},Numbering}\nonumber\\
&&\to \tt{Simple,ColumnsXRows} \to \{5, 2\}];
\eea
Here {\tt diagrams} is a variable used to store the output of {\tt InsertFields} and {\tt SMQCD} corresponds to the full SM. For convenience, one can also draw the generated diagrams with 
{\tt Paint}. The result of the evaluation in Eq.~\eqref{code5} is presented in Fig.~\ref{dsZSM}.
The option {\tt PaintLevel} is used to choose at which level ({\tt Classes}, in our example, with all up-type quarks, $u$, $c$, $t$ combined in $u_l$) the diagrams should be drawn. Similarly, the options {\tt Numbering} and {\tt ColumnsXRows} are used to format the picture \footnote{The options of {\tt Paint} are documented in {\tt FeynArts manual}\cite{famanual}.}. 

Then, we use the function {\tt PengCreateFeynAmp} to create amplitudes from {\tt diagrams}, as we see the role of it in Sec.\ref{structure}. 
\bea
\tt In[6]:=\hspace{.7cm}\tt{penguin = PengCreateFeynAmp[diagrams]}
\eea
Given the output of this command 
the function {\tt ExtractPenguinSubsRules} produces a list of substitution rules for each diagram (unevaluated amplitude) generated by {\tt PengCreateFeynAmp}. 
\bea
\tt In[7]:=\hspace{.7cm}\tt{substrules = ExtractPenguinSubsRules[penguin]}
\eea
The rules {\tt substrules} specify the actual couplings for each Generic diagram (amplitude). 
They can be used to obtain final analytic expressions in the considered model.
Given {\tt substrules}, one can obtain diagram-by-diagram contributions to the coefficient function of a particular operator by utilizing the {\tt SubstituteMassesAndFeynmanRules} function
\bea
\tt In[8]:=\hspace{.7cm}\tt{SubstituteMassesAndFeynmanRules[OP][substrules]}.
\eea
Here the string {\tt OP} can be chosen from the set {\tt\{"OpR", "OpL", "MonOpL0", "MonOpR0", "MonOpL2", "MonOpR2", "DipOpL1", "DipOpR1"\}}. The scalar operators from Eq.~\eqref{s1} correspond to {\tt "Op\{L,R\}"}, while monopole operators from Eq.~\eqref{e0ande2} correspond to {\tt"MonOp\{L,R\}\{0,2\}"} and finally, dipole operators from Eq.~\eqref{m1} --- to {\tt "DipOp\{L,R\}1"} . The output of {\tt In[8]} is a list of contributions to the considered Wilson coefficient for all diagrams given in Fig.\ref{gen}. For convenience, for every mass parameter {\tt M} we introduce a dimensionless mass ratio {\tt XXX[M/CommonMass]} with {\tt CommonMass} equals to the W-boson mass by default. The result is obtained in dimensional regularization. In spite of the fact that individual amplitudes
can have poles in the regularization parameter $\epsilon = (4-D)/2$, the sum is finite due to the absence of the tree-level FCNC.

\section{Test of {\tt Peng4BSM@LO}}
\label{test}
\begin{figure}[htb]
\begin{center}
\includegraphics[width=5in]{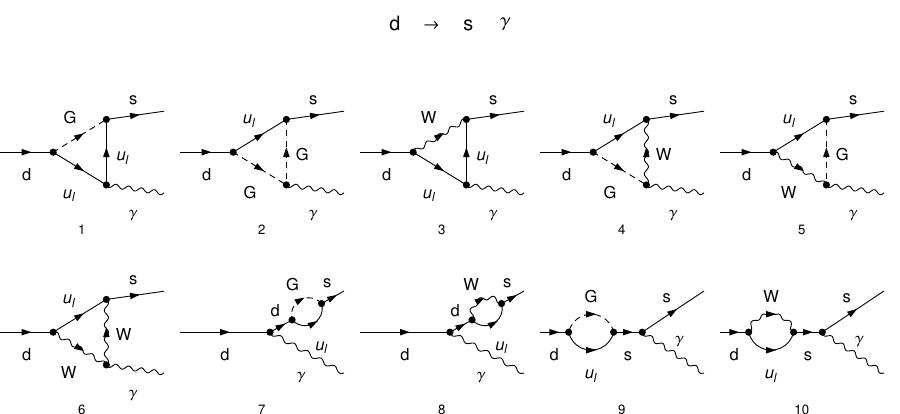}
\caption{The one-loop diagrams contributing to $d\bar s \gamma$ in the SM. The fields $u_l$, $W$, and $G$ in internal lines correspond to up-type quarks of the $l$-th generation,
$W$-boson and charged would-be Goldstone boson, respectively.}
\label{dsgammaSM}
\end{center}
\end{figure}

We compared\footnote{see {\tt Mathematica} notebook {\tt test\_Peng4BSMatLO.nb} included in the distribution.}  the file the results of application {\tt Peng4BSM@LO} to the induced $d\bar sZ$ vertex in the SM (see Fig.~\ref{dsZSM}) with those given in paper~\cite{inamilim} and got perfect agreement. For this particular case, the monopole operator is $\bar s_L \gamma^\mu d_LZ_\mu$ as in Eq.2.6 of~\cite{inamilim}. We also consider  the $\gamma$ exchange diagrams for the induced $d\bar s\gamma$ vertex in the SM (see Fig.~\ref{dsgammaSM}) for which the corresponding generic diagrams. The corresponding operators for the induced $d\bar s\gamma$ vertex are the monopole operators, $A_\mu\bar s\left(q^2\gamma_\mu-q_\mu\not q\right)P_Ld$ and the dipole operators, $A_\mu\bar s\sigma_{\mu\nu}iq^\nu\left(m_sP_L+m_dP_R\right)d$ as in Eq.B.1 of~\cite{inamilim}.
The consistency checks of {\tt Peng4BSM@LO} are as follows. There are no UV-divergencies in the considered form-factors. In the case of $Z$ boson only the left-handed FCNC operator is generated, i.e., $E^{0,Z}_R=0$. In the case of $\gamma$ quanta both $E^{0,\gamma}_{L,R} = 0$. As an example, we present the expression for $E^{0}_L$ corresponding to the monopole operator with external $Z$-boson:
\begin{eqnarray}
\label{e0le0r}
&&\left(E^{0,Z}_{c1,c2}\right)_{L}
=\sum_{k=2,3} \frac{e^3 V_{k1} V^*_{k2} \delta_{c1,c2} 
}{64\pi^2\cos\theta_W\sin^3\theta_W\left(x_1^2-1\right)^2\left(x_k^2-1\right)^2}\nonumber \\
&& \times  \Bigg\{\Big(x_1^2-1\Big)\Big(x_k^2-1\Big)\Big[x_k^2\Big(x_k^2-6\Big)+x_1^4\Big(x_k^2-1\Big)-x_1^2\Big(x_k^4-6\Big)\Big]\nonumber\\
&&+x_1^2\Big(6x_1^2+4\Big)\Big(x_k^2-1\Big)^2\log x_1-\Big(x_1^2-1\Big)^2x_k^2\Big(6x_k^2+4\Big)\log x_k\Bigg\}.
\end{eqnarray}
The expression is obtained by summing contributions from individual diagrams calculated by means of {\tt SubstituteMassesAndFeynmanRules} with {\tt OP="MonOpL0"}. In Eq.~\eqref{e0le0r} mass ratios $x_i=m_i/M_W$ are introduced with $i=1,2,3$ being up-type quark family indices. The Cabibbo-Kobayashi-Maskawa (CKM) matrix $V_{ij}$ is denoted in {\tt FeynArts} by $CKM(i,j)$. 
The Kronecker-delta in colour space is given by $\delta_{c1,c2}\equiv\textrm{IndexDelta}(c1,c2)$ with $c1$ and $c2$ being colour indices. The W-boson mass, the electric charge and sine of the Weinberg angle are denoted by $M_W$, $e$ and $\sin\theta_W$, respectively. 
It is worth pointing that it is crucial to use the unitarity of the CKM matrix to cancel divergent contributions to the form-factors~\eqref{e0le0r}. It is due to this,  only 
$V_{21} V^*_{22}\equiv V_{cd} V^*_{cs}  $ and $V_{31} V^*_{32}\equiv  V_{td}V^*_{ts}$ appear in Eq.~\eqref{e0le0r}.

The form factor for the second order monopole operator for a massless outgoing boson, in our case it is a photon $\gamma$, is:
\begin{eqnarray}
\label{e2le2r}
&&\left(E^{2,\gamma}_{c1,c2}\right)_{L}
=\sum_{k=2,3}\frac{e^3 V_{k1} V^*_{k2} \delta_{c1,c2}}{1152\pi^2M_W^2\sin^2\theta_W\left(x_1^2-1\right)^4\left(x_k^2-1\right)^4}\nonumber\\
&\times&\Bigg\{\Big(1-x_1^2\Big)\Big(x_k^2-1\Big)\Big[25x_k^4-19x_k^6+x_1^6\Big(19-57x_k^2+32x_k^4\Big)\nonumber\\
&+&x_1^4\Big(75x_k^2-32x_k^6-25\Big)+x_1^2\Big(57x_k^6-75x_k^4\Big)\Big]\nonumber\\
&+&4\Big(8-32x_1^2+54x_1^4-30x_1^6+3x_1^8\Big)\Big(x_k^2-1\Big)^4\log x_1\nonumber\\
&+&4\Big(1-x_1^2\Big)\Big(x_1^2-1\Big)^3\Big(8-32x_k^2+54x_k^4-30x_k^6+3x_k^8\Big)\log x_k\Bigg\}.
\end{eqnarray}
Finally,  a non-trivial contribution to the dipole operator for a massless photon is given by
\begin{eqnarray}
\label{m1lm1r}
&&\left(M^{1,\gamma}_{c1,c2}\right)_{L,R}
=-\sum_{k=2,3}\frac{i e^3 m_{s,d} V_{k1} V^*_{k2} \delta_{c1,c2}}{384 \pi^2 M_W^2 \sin^2\theta_W (x_1^2-1)^4 (x_k^2-1)^4}\nonumber\\
&\times&\Bigg\{\Big(x_1^2-1\Big) \Big(x_k^2-1\Big)\Bigg[x_1^6\Big(-29 x_k^4+31 x_k^2-8\Big)+x_1^4\Big(29 x_k^6-6 x_k^2-5\Big) \nonumber\\
&+&x_1^2\Big(-31 x_k^6+6 x_k^4+7\Big)+x_k^2 \Big(8 x_k^4+5 x_k^2-7\Big)\Bigg] \nonumber\\
&+&12 \Big(x_1^2-1\Big)^4x_k^4 \Big(3 x_k^2-2\Big)\log x_k-12 \Big(3 x_1^2-2\Big)x_1^4 \Big(x_k^2-1\Big)^4 \log x_1\Bigg\}
\end{eqnarray}
Also, we remind that in the presented equations the parameters defined in the {\tt SMQCD.mod} model file are used. It is obvious that the results can be rewritten in terms of Fermi constant $G_F=e^2/(8M^2_W\sin^2\theta_W)$ in order to factorize the result from the Wilson Coefficient together with the CKM matrix elements. 

\begin{figure}[htb]
\begin{center}
\includegraphics[width=3.5in]{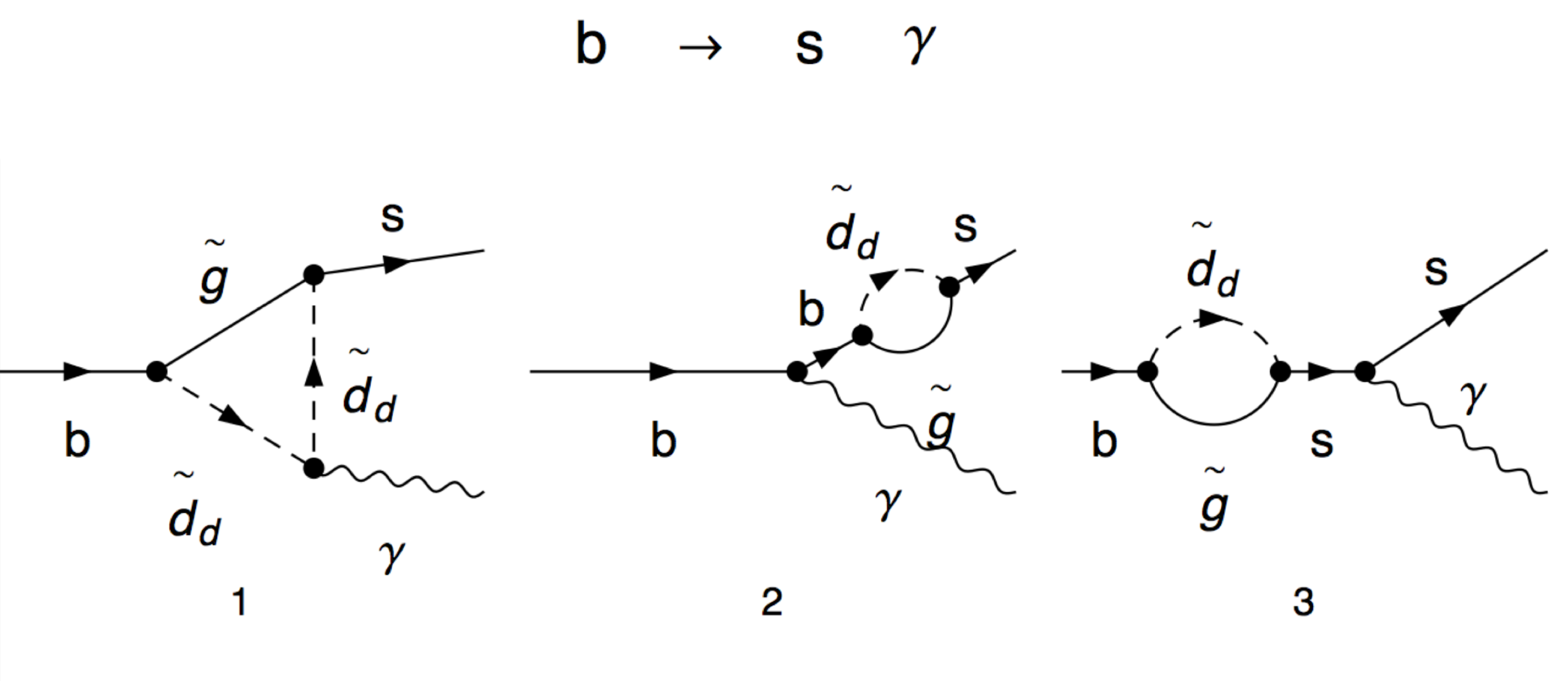}
\caption{The one-loop diagrams contributing to $b\bar s \gamma$ in the FVMSSM. Down-type scalar quarks
	and gluino are denoted by $\tilde{d}_d$ ($d=1,\dots,6$) and $\tilde g$, respectively.}
\label{bsgammaMSSM}
\end{center}
\end{figure} 

We also checked the correctness of ${\tt Peng4BSM@LO}$ in an application to the minimal supersymmetric standard model with non-minimal flavour violation (FVMSSM) \cite{Gabbiani:1996hi}.
The induced $b\bar s\gamma$ effective vertex in FVMSSM is given in Ref.~\cite{Bertolini:1990if}. The contribution due to gluino (see Fig.~\ref{bsgammaMSSM}) to the operator $im_b\epsilon_\mu\bar s\sigma^{\mu\nu}q_\nu P_Rb$ has the following form~\cite{Bertolini:1990if}:  
\begin{eqnarray}
\label{fvmssmforglu}
\mathcal{A}_{\tilde g}=-\frac{\alpha_s\sqrt{\alpha}}{\sqrt{\pi}}C(R)e_D\sum_{k=1}^6\frac{1}{m^2_{\tilde d_k}}\Bigg\{\Gamma_{DL}^{kb}\Gamma_{DL}^{\ast ks}F_2[x_k]-\Gamma_{DR}^{kb}\Gamma_{DL}^{\ast ks}\frac{m_{\tilde g}}{m_b}F_4[x_k]\Bigg\},
\end{eqnarray}
where $\alpha_s=g_s^2/4\pi$ is the strong coupling constant, $\alpha=e^2/4\pi$ corresponds to the fine structure constant, $x_k=m_{\tilde g}^2/m_{\tilde d_k}^2$ with $m_{\tilde g}$ being the mass of gluino and $m_{\tilde d_k}$ being the mass of the scalar quarks. The charge of down-type (s)quarks is $e_D=-1/3$, $\Gamma_{QL,R}$ are the $6\times 3$ squark mixing matrices, $C(R)=4/3$ is the quadratic Casimir operator on the fundamental representation of $SU(3): \sum_a(T^aT^a)_{ij}=C(R)\delta_{ij}$ with $Tr(T^aT^b)=\frac{1}{2}\delta^{ab}$ and
\begin{eqnarray}
F_2[x]&=&\frac{1}{12(x-1)^4}\left(2x^3+3x^2-6x+1-6x^2\log x\right)\nonumber\\
F_4[x]&=&\frac{1}{2(x-1)^3}\left(x^2-1-2x\log x\right)
\end{eqnarray}
The equation~\eqref{fvmssmforglu} should be compared with the expression produced by our package\footnote{The option {\tt ExcludeParticles->\{S[1|2|3|4|5|6],V[1|2|3|4|5],F[11|12]\}} was used to make {\tt PengInsertFields } generate diagrams given in Fig.~\ref{bsgammaMSSM}.}
for the operator $A_\mu\bar s\sigma^{\mu\nu}q_\nu P_Rb$:
\begin{eqnarray}
\label{packfvmssmforglu}
(M_{c1,c2}^{1,\gamma})_{R}&=&\frac{ieg_s^2}{288m_{\tilde g}^2\pi^2}\sum_{s=1}^6 \sum_{i=1}^3 \sum_{a=1}^8T^{a}_{c2,i}T^{a}_{i,c1}\nonumber\\
&\times&\frac{x_s}{(x_s-1)^4}\Bigg\{m_b\Bigg[1-6x_s+\Big(3-6\log x_s\Big)x_s^2+2x_s^3\Bigg]R_{\tilde d}^{\ast s_1,2}R_{\tilde d}^{s_1,3}\nonumber\\
&-&6m_{\tilde g}\Big(x_s-1\Big)\Big(-1-2x_s\log x_s+x_s^2\Big)R_{\tilde d}^{\ast s_1,2}R_{\tilde d}^{s_1,6}\Bigg\}.
\end{eqnarray}
Here $R^{s_1,s_2}_{\tilde{d}}$ is the $6\times 6$ down-type squark mixing matrix such that $R_{\tilde d}=\left(\Gamma_{DL}|\Gamma_{DR}\right)$. The generators of SU(3) are denoted by $T^a_{ij}$. 
In Eq.~\eqref{packfvmssmforglu} we neglect explicit dependence on $m_s$. It is interesting to note that in the same approximation the coefficient $(M_{c1,c2}^{1,\gamma})_{L}$ of the operator $A_\mu\bar s\sigma^{\mu\nu}q_\nu P_L b$ can be obtained
from Eq.~\eqref{packfvmssmforglu} by the substitutions $R^{s_1,2}_{\tilde d} \leftrightarrow R^{s_1,5}_{\tilde d}$  and $R^{s_1,3}_{\tilde d} \leftrightarrow R^{s_1,6}_{\tilde d}$, which correspond to $\Gamma_{DL}\leftrightarrow \Gamma_{DR}$ replacement in Eq.~\eqref{fvmssmforglu}. 
After some obvious colour algebra one can see that $i m_b\mathcal{A}_{\tilde g}\delta_{c1,c2}=(M_{c1,c2}^{1,\gamma})_{R}$.

In Eq.~\eqref{packfvmssmforglu} obtained by {\tt Peng4BSM@LO}, the parameters are as in the model file, {\tt FVMSSM.mod}, located in the Models subdirectory of {\tt FeynArts}, {\it i.e.} the corresponding definitions of the parameters in {\tt FVMSSM.mod} as the following: $m_b=\textrm{MB}$, $g_s=\textrm{GS}$, $e=\textrm{EL}$, $m_{\tilde g}=\textrm{MGl}$ and $R^{s_1,s_2}_{u,d}=\textrm{UASf[t][$s_1$,$s_2$]}$, $T^a_{ij}=\textrm{SUNT[a,i,j]}$ and the sums $\sum_{i=1}^r$ are represented by the factors $\textrm{SumOver[i,r]}$.

\begin{figure}[htb]
\begin{center}
\includegraphics[width=5in]{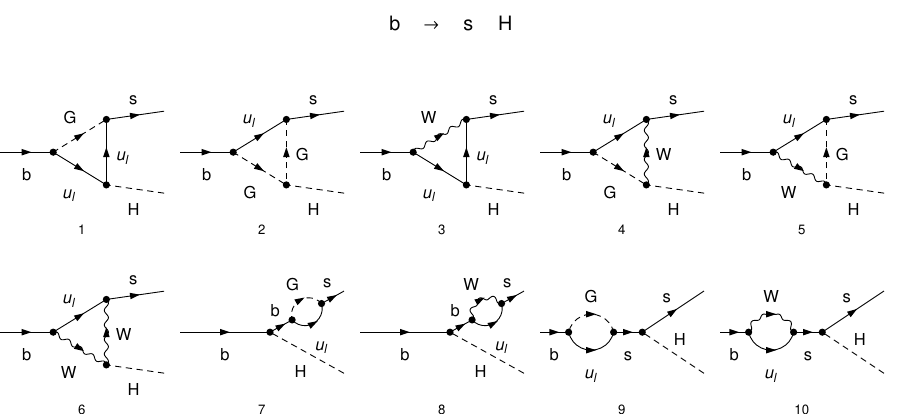}
\caption{The one-loop diagrams contributing to $b\bar sH$ in the SM.}
\label{bsHSM}
\end{center}
\end{figure}

The third example of the application of {\tt Peng4BSM@LO} is to the SM Higgs penguin with quark flavour changing interactions. For the effective Lagrangian~\cite{Dedes:2003kp}
\begin{eqnarray}
\label{lagforH}
\mathcal L^{SM}_{H\bar sb}= -\frac{g_2}{2M_W}H\bar s\bigg[ m_s 
\left(\bold g^L_{H\bar dd^\prime}\right)_{sb} P_L
+ m_b\left(\bold g^R_{H\bar dd^\prime}\right)_{sb} P_R\bigg]b~,
\end{eqnarray}
the induced $b\bar s H$ effective vertex in the SM (see Fig.~\ref{bsHSM}) is given in \cite{Dedes:2003kp}. The contribution corresponding to the operators $m_{s,b}H\bar  sP_{L,R}b$ can be inferred from the matrix elements $\left(\bold g^{L,R}_{H\bar dd^\prime}\right)_{sb}$ of the factors \cite{Dedes:2003kp}:
\begin{eqnarray}
\label{higgscont}
\bold g^L_{H\bar dd^\prime}=-\frac{g_2^2}{(16 \pi^2)}\bold V^\dagger f(\hat x,y)\bold V~~~,~~\bold g_{H_i\bar dd^\prime}^R=\left(\bold g_{H_i\bar dd^\prime}^L\right)^\dagger
\end{eqnarray}
where $\hat x=\bold{\hat M}_u^2/M_W^2$, $y=M_H^2/M_W^2$ and
\begin{eqnarray}
f(\hat x,y)=\frac{3}{4}\hat x+y\bigg(-\frac{\hat x^3\ln\hat x}{4(1-\hat x)^3}+\frac{\hat x^2\ln\hat x}{2(1-\hat x)^3}-\frac{\hat x^2}{8(1-\hat x)^2}+\frac{3\hat x}{8(1-\hat x)^2}\bigg)
\end{eqnarray}
Hereby the up- and down-type $3\times 3$ \emph{diagonal} quark mass matrices are denoted by $\bold{\hat M}_u$ and $\bold{\hat M}_d$, respectively, $M_H$ is the Higgs boson mass and $\bold V$ corresponds to the CKM matrix.

Taking into account  Eq.~\eqref{higgscont}, the expression in Eq.~\eqref{lagforH} can be compared with the result produced by the package {\tt Peng4BSM@LO} for the effective $b\bar s H$ vertex. In the limit $m_u = m_c = 0$ one has
\begin{eqnarray}
\label{scalarbypackLR}
&&(N^{0,H}_{c_1,c_2})_{L,R} =  \frac{e^3m_{s,b}V_{33} V^*_{32}\delta_{c1,c2}}{256M_W\pi^2\sin^3\theta_W}\nonumber\\
&& \times  \bigg\{6x_3+\frac{y}{(-1+x_3)^3}\bigg[-3x_3+4x_3^2-4x_3^2\ln x_3-x_3^3+2x_3^3\ln x_3\bigg]\bigg\}
\end{eqnarray}
where $x_3=m_t^2/M_W^2$, $y=M_H^2/M_W^2$ and $e=g_2\sin\theta_W$. The equivalence between the results of Ref.~\cite{Dedes:2003kp} and the output of the package  is obvious in the considered limit: 
$-\frac{g_2}{2 M_W} m_{s,b}(\delta_{c_1,c_2}. g^{L,R}_{H\bar sb})=(N^{0,H}_{c_1,c_2})_{L,R}$. The comparison of Eqs.~\eqref{higgscont}~and~\eqref{scalarbypackLR} is affirmative for   {\tt Peng4BSM@LO}. 

In addition, we would like to mention that our package correctly reproduces the general results for $f_1 f_2 \gamma$ vertex~\cite{Lavoura:2003xp} in QED with additional scalar boson.

\section{Conclusion}
We present the  new package which we called {\tt Peng4BSM@LO}. {\tt Peng4BSM@LO} is written in {\tt Mathematica} and works with {\tt FeynArts} and/or {\tt FeynCalc}. The package  defines and calculates contributions to the Wilson coefficients of particular operators for the one-loop penguin diagrams in FCNC processes. \footnote{Hereby, it should be mentioned that recently an extension of {\tt SARAH}, {\tt FlavorKit}\cite{Porod:2014xia}, which handles flavour observables, became available. The authors of {\tt FlavorKit} also utilized our package to cross check some of their results.}

We conducted thorough testing of the package and reproduced known results for the induced $d\bar sZ$ and $d\bar s\gamma$ vertices  in the SM~\cite{inamilim}, for the gluino contribution to $b\bar s\gamma$ in the MSSM with non-minimal flavour violation~\cite{Bertolini:1990if}, and for the induced $b\bar sH$ vertex in the SM~\cite{Dedes:2003kp}. This serves as a validity check of our code.

The advantage of the package is that it relies on the general Lorentz structure of the penguin amplitudes and, as a consequence, can be used to evaluate the Wilson coefficients in any renormalizable model which extends the SM.

The next steps are the calculation of the box diagrams and implementation of the Flavour Les Houches Accord \cite{Mahmoudi:2010iz} output, which allows one to carry out a full calculation of flavour observables.

\section*{Acknowledgments}
We thank D.I. Kazakov for reading the manuscript and providing us with valuable comments and suggestions. This work is partially supported by RFBR (Russian Foundation for Basic Research) grant No. 14-02-00494-a. 
Additional support from JINR Grant No. 14-302-02 and Dynasty Foundation is kindly acknowledged by AVB.

\appendix

\section{Program Summary}
\label{summary}

\begin{small}
\begin{itemize}

\item{\it Title of program:}
  {The name of {\tt Peng4BSM@LO} is the abbreviation of penguin diagrams for Beyond the Standard Model (SM) in the leading order.}
  
  \item{\it Available from:}
  \url{http://theor.jinr.ru/~hanif/peng4bsm@lo/}
  
  \item{\it Programming Language:} {\tt Mathematica}

  \item{\it Computer:} Any computer where {\tt Mathematica 8} and newer is running. 
  
  \item{\it Operating system:} {\tt Windows}, {\tt Linux}, {\tt MacOSX}.  

  \item{\it Number of bytes in distributed program including test data etc.: }
  	  {\tt Peng4BSMatLO.m} is $\sim$1 065 457 bytes, {\tt test\_Peng4BSMatLO.nb} is $\sim$673 655 bytes.
  
  \item{\it Distribution format:} 
  ASCII
  \item{\it External routines/libraries:} {\tt FeynArts 3.7} or {\tt FeynCalc 8.2}

\item{\it Keywords:} Penguin, BSM, Beyond the Standard Model, the SM, the Standard Model, FCNC, Wilson Coefficients, Effective Hamiltonian, OPE, Operator Product Expansion, CP violation.

\item{\it Nature of physical problem:}
  FCNC processes are absent in the SM at the tree level and, thus, are suppressed in comparison to the charged current processes. Due to this, FCNC can be used as 
  an excellent probe of New Physics which can considerably alter the predictions of the SM.  
The rare processes impose very important constraints on Beyond-the-SM (BSM) physics.

\item{\it Method of solution:}
	{\tt Peng4BSM@LO} uses {\tt Mathematica} to evaluate relevant contributions to the considered Wilson coefficients from penguin diagrams in the SM 
	and Models Beyond the SM.
  
\item{\it Restrictions:}
  The calculations are restricted to the case of renormalizable Quantum Field Theories quantized in Feynman gauge. The standard generic model file, {\tt Lorentz.gen}, distributed with {\tt FeynArts} should be used.  

\item{\it Typical running time:}
  For all operations the running time does not exceed $\sim$60 seconds for the SM and the MSSM which we have chosen for testing. However, the running time significantly depends on the number of parameters and fields of the chosen model.
  
\end{itemize}
\end{small}

\section{Description of the Main Procedures}
\label{procedure}

\begin{itemize}

\item {\tt PengInsertFields[\{InF\} $\to$ \{OutF, OutV\}, Model -> MOD]}
	
\textit{General}: Equivalent to the {\tt FeynArts} function {\tt InsertFields} and is used to construct all Feynman penguin-type diagrams in a particular model for a particular set of external fields from a predefined set of topologies {\tt PenguinTopologies}.

\textit{Input}: {\tt InF}=$F_a$, {\tt OutF}=$F'_b$, and {\tt OutV}=$V_c$ specify external fermions and vector fields (see Eqs.~\eqref{e0ande2}-\eqref{m1}), which are used to construct the corresponding induced operator, {\tt MOD} is a {\tt FeynArts} model. Accepts the same options as {\tt InsertFields}.

\textit{Output}: {\tt TopologyList[\ldots]} --- a hierarchical\footnote{Reflects the {\tt FeynArts} hierarchy {\tt Generic} - {\tt Classes} - {\tt Particles}.} list of penguin-type Feynman diagrams in {\tt FeynArts} notation.

\item {\tt PengCreateFeynAmp[ diagrams ]}

	\textit{General}: Similar to the {\tt FeynArts} function {\tt CreateFeynAmp}, which produces analytic expressions for amplitudes given the diagrams {\tt diagrams} created with the help of {\tt InsertFields}.

	\textit{Input}: {\tt diagrams} ---  diagrams created by {\tt PengInsertFields}. Accepts the same options as {\tt CreateFeynAmp}

	\textit{Output}: {\tt PengFeynAmpList[\ldots][\ldots]} --- a list of analytic expressions for {\tt Generic} amplitudes together with the required substitution rules.

\item {\tt ExtractPenguinSubsRules[ penguins ]}

\textit{General}: Extracts a list of substitution rules for each amplitude from the output of {\tt PengCreateFeynAmp}.
			   The rules specify the actual couplings for each Generic diagram (amplitude). In addition, information about the required Generic
			   diagram is stored. 
			   
\textit{Input}: {\tt penguins} --- the output of {\tt PengCreateFeynAmp}.

\textit{Output}: {\tt PenguinSubsRules[\ldots][\ldots]} ---  a list of substitution rules for each diagram (amplitude) generated by {\tt PengCreateFeynAmp}.

\item {\tt SubstituteMassesAndFeynmanRules}[{\tt OP}][ {\tt substrules} ]

	\textit{General}: Applies the substitution rules to the predefined {\tt Generic} coefficient function specified by tag {\tt OP}, given the rules generated by {\tt ExtractPenguinSubsRules}.

	\textit{Input}: {\tt OP = \{ "OpL" | "OpR" | "MonOpL0" | "MonOpR0" | "MonOpL2" | "MonOpR2" | "DipOpL1" | "DipOpR1" \}} --- the operator type, 	
	 {\tt substrules} ---  the output of  {\tt ExtractPenguinSubsRules}.

\textit{Output}: A list with diagram-by-diagram contributions to the coefficient functions of the specified operator {\tt OP}.

\end{itemize}

\section{Description of the Auxiliary Procedures and Definitions}
\label{auxiliary}

\begin{itemize}

\item {\tt \$UseFeynCalc }
	
	\textit{General}: Controls whether {\tt FeynCalc} should be used ({\tt \$UseFeynCalc = True}) 
	in place of {\tt FeynArts.}

\item {\tt eps }
	
	\textit{General}: Parameter of dimensional regularization $D=4 - 2 \epsilon$.

\item {\tt CommonMass }
	
	\textit{General}: A mass which is used to form dimensionless ratios, {\tt CommonMass = $M_W$ = MW} by default. It can be redefined for the user convenience.

\item {\tt UnitarityCKM[ V, Ngen ] }
	
\textit{General}: Generates a list of substitution rules for non-diagonal matrix elements of the CKM matrix which reflect the unitarity of the latter.

\textit{Input}: {\tt V} --- the name of the CKM matrix as defined in the considered model (e.g.,{\tt CKM} in "SMQCD"), {\tt Ngen} = $n_g$ --- number of fermion generations

\textit{Output}: A list of rules similar to $V_{i1} V^*_{j1} \to - \sum\limits_{k=2}^{n_g} V_{ik} V^*_{jk}$.

\item {\tt CollectSumOver[ expression ] }
	
	\textit{General}: Converts recursively the expressions involving sums over different indices in {\tt FeynArts} notation (e.g., {\tt (a[i]*SumOver[i,1,N] + b[i]*SumOver[i,1,N] + \ldots)}) to new notation {\tt IndexSum[ a[i] + b[i] + .., \{i,1,N\}]}. 

	\textit{Input}: an expression containing {\tt FeynArts} sums with {\tt SumOver}.

	\textit{Output}: the same expression rewritten in terms of {\tt IndexSum}.

\item {\tt ExpandInSmallMasses[ expression, {masslist}, order ] }
	
	\textit{General}: Expands the given expression in small masses up to the given order. 

	\textit{Input}: {\tt expression} --- an expression to be expanded, {\tt masslist = \{ m1, m2, \ldots\}} --- a list of masses which are assumed to be small, {\tt order} --- all the terms of the order of ({\tt order + 1}) will be neglected in the output.

	\textit{Output}: the expanded expression.

\item {\tt XXX[Mass/CommonMass] }
	
	\textit{General}: The output of {\tt SubstituteMassesAndFeynmanRules} 
	is written in terms of dimensionless mass ratios {\tt XXX[Mass/CommonMass]} and a common mass {\tt CommonMass}.

\end{itemize}


\end{document}